\documentclass[%
 reprint,
groupedaddress,
 amsmath,amssymb,
 aps,
]{revtex4-1}

\usepackage{graphicx}
\usepackage{dcolumn}
\usepackage{bm}

\begin{document}


\title{Study of planar Ising ferromagnet on the triangular lattice with selective dilution}

\author{T.~Balcerzak}
\email{t\_balcerzak@uni.lodz.pl; kszalowski@uni.lodz.pl}
\author{K.~Sza{\l}owski}
\email{t\_balcerzak@uni.lodz.pl; kszalowski@uni.lodz.pl}
\affiliation{Department of Solid State Physics, University of
\L\'{o}d\'{z}\\Pomorska 149/153, PL-90-236 \L\'{o}d\'{z}, Poland}
\author{M. \v{Z}ukovi\v{c}}
\author{M. Borovsk\'{y}}
\author{A. Bob\'{a}k}
\author{M. Ja\v{s}\v{c}ur}
\affiliation{Department of Theoretical Physics and Astrophysics\\ Faculty of
Science, P. J. \v{S}af\'arik University\\ Park Angelinum 9, SK-041 54
Ko\v{s}ice, Slovak Republic}

\date{\today}

\begin{abstract}
In the paper the Curie temperatures of selectively diluted
planar Ising ferromagnet on the triangular lattice are calculated vs.
concentration of magnetic atoms. Various analytical approaches are compared
with the exact numerical calculations for finite clusters, as well as with
the exact analytical solutions for the triangular and honeycomb lattices.

\end{abstract}

\maketitle


\section{\label{sec:level1}Introduction}

The studies of low-dimensional magnets have presented a topical item for
many years. The investigations include both 1D and 2D magnets, as well as
bilayers, thin films and multilayers. Many different theoretical methods
have been employed; in some cases [1--6] the exact solutions are available.

As far as approximate methods are concerned, one should mention the Green
Function (GF) [7--9] and spectral density [10] methods, Renormalization Group (RG) approach [11, 12], including Mean-Field Renormalization Group (MFRG) [13, 14] and Effective-Field Renormalization Group (EFRG) [15],
Spin-Wave (SW) techniques [16--18], High Temperature Series Expansion
(HTSE) [19] and Monte Carlo (MC) simulations [20, 21]. 
Many other approaches like Coherent Anomaly Method (CAM) together with transfer matrix technique [22] and Effective Field Methods (EFM) with correlations [23] should also be mentioned.
Recently, Cluster Variational Method in the Pair Approximation (PA) has been adopted for studies of the bilayer [24] and bi-multilayer [25] systems.

The aim of the present paper is to study the ferromagnetic Ising model
with spin $S=1/2$ on the Planar Triangular (PT) lattice with selective
dilution. By the selective dilution we mean the dilution of only one
sublattice, whereas the PT magnet can be, in general, decomposed into three
interpenetrating sublattices. Regarding antiferromagnetism and the problem
of frustration, such selectively diluted model has been considered by Kaya
and Berker [26]. To the best of our knowledge, as far as ferromagnetism is
concerned, the model has not been studied yet.

The model is interesting from the theoretical point of view, for by
changing the selective dilution parameter we are able to pass continuously
from the ideal PT lattice (without dilution) to the honeycomb lattice,
where one sublattice is completely empty. On the other hand, for those two
cases the exact solutions for the Curie temperatures do exist [2]. In this
paper we will concentrate on the Curie temperature calculations for
arbitrary concentration of magnetic atoms in the selectively diluted
sublattice. Thus, we consider an intermediate situation between those two
limiting cases, which were examined exactly by Wannier [2].

In the next Section, the outline of the theoretical methods will be given,
and the respective formulas for the phase transition temperatures will be
presented. With the help of numerical calculations we are able to compare
the results of several theoretical approaches, namely the Molecular Field
Approximation (MFA), Effective Field Theory (EFT), Pair Approximation (PA)
method, as well as the Exact Calculation for Finite Clusters (ECFC). The
results will be presented in the plots and discussed.

\section{Theoretical methods}

Let us consider the ferromagnetic Ising model with spin $S=1/2$ on the PT
lattice with selective dilution. The diluted lattice is illustrated in
Fig.\,1. Following Kaya and Berker [26] we keep decomposition of the PT
lattice into three interpenetrating sublattices $a$, $b$ and $c$, of which,
for the ferromagnetic case the sublattices $a$ and $b$ will be equivalent.
The sublattice $c$ is distinguished in the system by the random dilution of
the spins. If we denote by $p$ the concentration of spins on
$c$-sublattice, the case of $p=0$ corresponds to the honeycomb lattice,
whereas $p=1$ stands for the non-diluted triangular lattice.

The Hamiltonian of the system is in the form of:

\begin{equation}
\label{eq0}
{\cal H} =- J\sum \limits_{\left( i_a,j_b\right) } S_{i_a}^{z}S_{j_b}^{z}
- J\sum \limits_{\left( i_a,j_c\right) } S_{i_a}^{z}S_{j_c}^{z}\xi_{j_c}
- J\sum \limits_{\left( i_b,j_c\right) } S_{i_b}^{z}S_{j_c}^{z}\xi_{j_c}
\end{equation}

\noindent where $J>0$ is the exchange interaction coupling, and $\left(
i_\alpha,j_\beta \right)$ means the summation extending over nearest
neighbour pairs of spins from sublattice $\alpha$ and $\beta$. $\xi_{j_c}$
are quenched, uncorrelated random variables chosen to be equal 1 with
probability $p$ when the site $j_c$ is ocupied by a magnetic atom and 0
with probability $1-p$ otherwise.

\begin{figure}[h]
\includegraphics[scale=0.87]{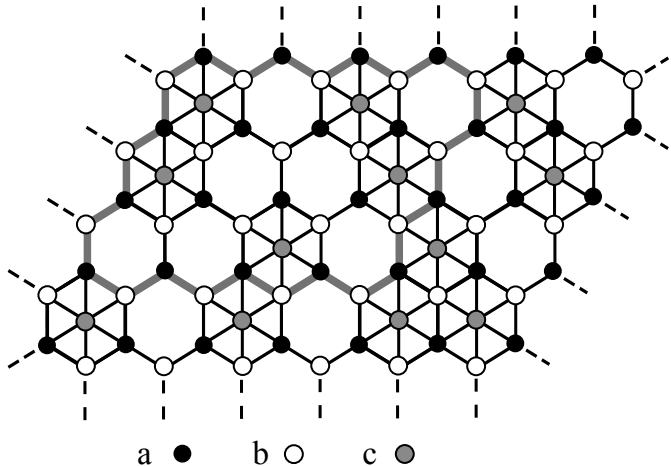}
\caption{The triangular lattice divided into three interpenetrating sublattices:
$a$, $b$ and $c$. The sublattice $c$ is randomly diluted. The thick grey line
surrounds a cluster containing $3\times 4$ hexagons, used for exact calculations.}
\label{fig1}
\end{figure}

\subsection{Molecular Field Approximation (MFA)}

In this simplest approach the sublattice magnetizations $m_a=\left<
S_{i_a}^{z}\right>$, $m_b=\left< S_{i_b}^{z}\right>$ and $m_c=\left<
S_{i_c}^{z}\right>$ are described by three coupled MFA equations. In the
vicinity of the Curie temperature the equations can be linearized and
presented as follows:
{\setlength\arraycolsep{1pt}
\begin{eqnarray}
\label{eq1}
m_a& = &\frac{3}{4}\beta_{\rm C}J\left(m_b+m_c\,p\right)\nonumber \\
m_b &= &\frac{3}{4}\beta_{\rm C}J\left(m_a+m_c\,p\right)\nonumber\\
m_c &= &\frac{3}{4}\beta_{\rm C}J\left(m_a+m_b\right)
\end{eqnarray} }

\noindent where  $\beta_{\rm C}=1/k_{\rm B} T_{\rm C}$,  and $T_{\rm C}$
denotes the Curie temperature.

It is convenient to introduce the variable $t_{\rm C}=k_{\rm B} T_{\rm
C}/J$  which is a dimensionless Curie temperature. Then, assuming symmetry
condition for the ferromagnet $m_a=m_b\ne m_c$, and demanding that the
determinant of eqs.\,(2) must be zero, we obtain the equation for the Curie
temperature in MFA:
\begin{equation}
\label{eq2}
8 t_{\rm C}^2 - 6 t_{\rm C}- 9p=0.
\end{equation}
The physical solution is then of the form:
\begin{equation}
\label{eq3}
t_{\rm C}=\frac{3}{8}\left(1+\sqrt{1+8p}\right)
\end{equation}
and is straightforward for numerical calculation.
 \newline

\subsection{Effective Field Theory (EFT)}
By the EFT we mean the method proposed by Honmura and Kaneyoshi [27],
which takes into account autocorrelations but neglects the spin-pair
correlations. Among its many applications, the method has recently been
applied for the triangular lattice with uniform dilution [28]. It is based
on the exact Callen-Suzuki identity of the form:
\begin{equation}
\label{eq4}
\left<S_{i_\alpha}^z\right>=\frac{1}{2}\left<\tanh \frac{1}{2k_{\rm B}T}
\left(J\sum_{j_{\beta} \in i_{\alpha}}S_{j_\beta}^z\xi_{j_\beta}\right)\right>.
\end{equation}

\noindent where $\xi_{j_\beta}=1$, if $\beta=a,b$ and $j_{\beta}\in
i_{\alpha}$ denotes a lattice site being nearest neighbour of the site
$i_{\alpha}$ .

Applying the differential operator method [27], together with the
decoupling procedure for the mean value of multi-spin products, the local
magnetizations can be calculated. The coupled equations for the sublattice
magnetizations have polynomial form and can be linearized near the
continuous phase transition points. For the system in question, with the
general sublattice magnetizations $m_a$, $m_b$ and $m_c$, the Curie
temperature can be found from the determinant equation:
\begin{equation}
\label{eq5}
	\det {\bf U} = 0,
\end{equation}

\noindent where
\begin{widetext}
\begin{equation}
\label{eq6}
{\bf U} = \left(
\begin{array}{ccc}
-1 & 3 a^2 a_{c}^3 b \tanh \left( x \right) |_{x=0} & 3 a^3 a_{c}^2 b \tanh \left( x \right) |_{x=0} \\
3 a^2 a_{c}^3 b \tanh \left( x \right) |_{x=0} & -1 & 3 a^3 a_{c}^2 b \tanh \left( x \right) |_{x=0} \\
3 p a^5 b \tanh \left( x \right) |_{x=0} & 3 p a^5 b \tanh \left( x \right) |_{x=0} & -1 \\
\end{array}
\right).
\end{equation}
\end{widetext}

\noindent In eq.\,(7) the temperature-dependent coefficients are given in
the following form:
\begin{eqnarray} \label{eq.7} a &=& \cosh \left(
\frac{1}{4t_{\rm C}} D \right) \nonumber\\ a_{c} &=& 1 - p + p \cosh \left(
\frac{1}{4t_{\rm C}} D \right)\nonumber\\ b &=& \sinh \left( \frac{1}{4t_{\rm
C}} D \right), \end{eqnarray}

\noindent where $D=\partial/\partial x$ is the differential operator. The
equation\,(\ref{eq5}) can be solved numerically only.

\subsection{The Pair Approximation (PA) method}

The PA is one of the cluster variational methods and takes into account
the nearest-neighbour correlations. Being more accurate than MFA and EFT it
enables calculation of the Gibbs energy, and hence all thermodynamic
properties. Contrary to MFA, in the PA method the local variational
parameters (molecular fields acting on a pair) are no longer simply
proportional to local magnetizations. A set of linearized equations for
these parameters near the Curie temperature takes a form:
\begin{eqnarray}
\label{eq.8}
\left(3C_1-2\right)\lambda + \left(3C_1-3\right)p\lambda_2&=&0\nonumber\\
\left(3C_1-3\right)\lambda + \left(6C_1-5\right)\lambda_1 + \left(3C_1p-3p+1\right)\lambda_2&=&0\nonumber\\
\left(3C_2-3\right)\lambda - \left(6C_2-5\right)\lambda_1 + \left(3C_2p-3p+1\right)\lambda_2&=&0.\nonumber\\
\end{eqnarray}

\noindent The variational parameters have the following meaning:\\
$\lambda$ is the field acting on a spin on the sublattice $a$ or $b$ and originating from spins
on the sublattices $b$ or $a$, respectively;\\
$\lambda_1$ is the field acting on the spin on the sublattice $c$
and originating from the sublattices $a$ or $b$;\\
$\lambda_2$ is the field acting on the spin on the sublattice $a$ or $b$ and originating
from the sublattice $c$.

The temperature-dependent coefficients have the following form:
\begin{eqnarray}
\label{eq.9}
C_1&=&\frac{1}{2}\left(1+\frac{1}{x_{\rm C}}\right)\nonumber\\
C_2&=&\frac{1}{2}\left(1+x_{\rm C}\right)
\end{eqnarray}

\noindent where
\begin{equation}
\label{eq.10}
x_{\rm C}=\exp \left(\frac{1}{2t_{\rm C}}\right).
\end{equation}

\noindent By setting the determinant of eqs.\,(9) equal to zero the Curie
temperature can be found. The final result can be presented in the form of
the algebraic equation:
\begin{equation} \label{eq.11}
3\left(1+5p\right)x_{\rm C}^3 -\left(13+30p\right)x_{\rm C}^2
+15\left(1+p\right)x_{\rm C} -9=0 \end{equation}

\noindent where $x_{\rm C}$ is
related to the Curie temperature, $t_{\rm C}=k_{\rm B} T_{\rm C}/J$, by the
formula (11).

\subsection{Exact Calculation for Finite Clusters (ECFC)}

Exact numerical diagonalization for finite systems is nowadays a powerful
tool for the studies of magnetic properties [29]. The accuracy of this
method improves with the increase of the cluster size, but simultaneously
rapidly growing number of states, which should be taken into account,
results in a corresponding huge increase of the calculation time. This
requires increasingly powerful computers. However, in the case of spin systems with Ising interactions, all the system states and their energies can be listed explicitly, without resorting to diagonalization of the Hamiltonian. Hence, we call this approach Exact Calculation for Finite Clusters. The method bears some resemblance to Monte Carlo calculations, however, it uses all the system states to study its thermodynamics within canonical ensemble. The method is sensitive to the
shape of a cluster and selection of the boundary conditions. It is known from the literature devoted to Monte Carlo simulations that the selection of periodic boundary conditions is evaluated as superior to other choices for planar lattices [30,31]. In particular, it guarantees the same number of nearest-neighbours for the atoms on the boundary and inside the cluster. In our calculations presented here, we based on a cluster consisting of $3 \times 4$ hexagons, and the boundary conditions
were chosen as periodic. Such a cluster is illustrated in
Fig.\,\ref{fig1}, where it is surrounded by the gray thick line. The maximum
number of spins (for $p=1$, when all sites on the sublattice $c$ were
occupied) amounted to 36, while for $p=0$ it was equal to 24. 

The Curie temperatures were determined from the maxima of the specific
heat curves for various concentrations $p$. The condition for the maximum
can be found from the exact thermodynamic formula:
\begin{equation}
\label{eq.12}
\left<E^3\right>-3\left<E^2\right>\left<E\right>+2\left<E\right>^3=
2\left[\left<E^2\right>-\left<E\right>^2\right]k_{\rm B}T_{\rm max},
\end{equation}

\noindent where the mean values of the energy powers $E^n$ are calculated
numerically with the Boltzmann distribution (taken at the temperature
$T_{\rm max}$) over all possible states in the cluster. For our purpose, in
eq.\,(13) $T_{\rm max}$ for a finite cluster is assumed to estimate $T_{\rm C}$, which is a common approach used in Monte Carlo studies, e.g. [32,33]. According to scaling relations, $T_{\rm max}$ is expected to converge to $T_{\rm C}$ in the limit of an infinite system size [33,34]. In the case of ECFC, the system size is severely limited by the computational resources and thus extrapolation to infinite system size would be not trustworthy, due to small-size corrections to scaling. Therefore, we present directly the obtained values of $T_{\rm max}$ for the largest system studied, i.e. the 3$\times$4 cluster and for comparison we provide also the numbers for smaller 3$\times$3 cluster. 

The numerical results are presented in the next Section.

\section{Numerical results and discussion}

The Curie temperatures for selectively diluted PT ferromagnet have been
calculated based on the approximations presented in the previous Section.
The results of $k_{\rm B} T_{\rm C}/J$ vs. concentration $p$ are
illustrated in Fig.\,2. In the same figure two exact Wannier results [2]
are shown, i.e., for $p=0$ (honeycomb lattice) and for $p=1$ (triangular
lattice). For the pure honeycomb lattice we obtained the Curie temperature
values $k_{\rm B} T_{\rm C}/J$ equal to: 3/4 (MFA); 0.5259 (EFT);
$1/\left(2\ln3\right)\approx$0.4551 (PA); 0.4203 (ECFC for 3$\times$3 cluster) and 0.4128 (ECFC for 3$\times$4 cluster). These results
can be compared with the exact Wannier solution $k_{\rm B} T_{\rm
C}/J=1/\left(2\ln(2+\sqrt{3})\right)\approx$0.3797. On the other hand, for
the pure triangular lattice the Curie temperatures calculated in various
approaches are: $k_{\rm B} T_{\rm C}/J$=3/2 (MFA); 1.2683 (EFT);
$1/\left(2\ln(3/2)\right)\approx$1.2332 (PA); 0.9602 (ECFC for 3$\times$3 cluster) and 0.9520 (ECFC for 3$\times$4 cluster). The exact
Wannier result in this case is $k_{\rm B} T_{\rm
C}/J=1/\ln3\approx$0.9102.
\begin{figure}[t]
\begin{center}
\includegraphics[scale=0.63]{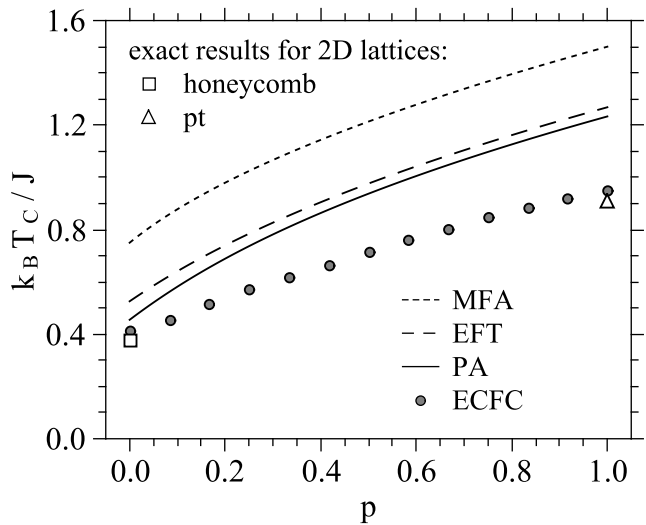}
\caption{The Curie temperatures vs. concentration for the triangular lattice
with selective dilution, obtained by various theoretical approaches: MFA, EFT,
PA and ECFC for 3$\times$4 cluster. Two exact Wannier results [2] for $p=0$ and $p=1$ are also indicated.}
\label{fig2}
\end{center}
\end{figure}
\begin{figure}[h]

\vspace{5mm}

\begin{center}
\includegraphics[scale=0.63]{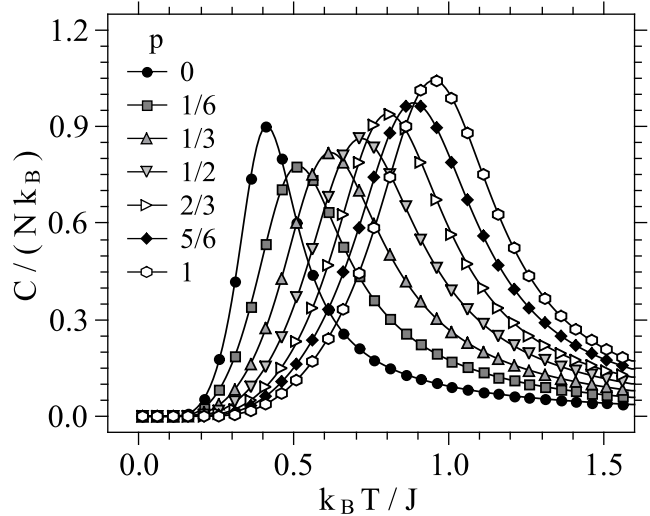}
\caption{The specific heat per lattice site vs. temperature for various
concentrations of $c$-atoms. The presented results are obtained by the
ECFC method for 3$\times$4 cluster.}
\label{fig3}
\end{center}
\end{figure}

The Curie temperature of the pure triangular lattice is higher than that
of the honeycomb one, since the coordination number of the former doubles
that of the latter. It is seen in Fig.\,2 that in each method the Curie
temperature changes continuously with $p$;  however, the change is not
linear, not even in the MFA.

Fig.\,2 illustrates the accuracy of the methods described in the previous
Section. As far as the analytical methods are concerned, we see that MFA is
the least accurate; giving the highest Curie temperature. The EFT and PA
are much more accurate methods. Noticeably, the numerical calculations
performed on the finite clusters with periodic boundary conditions seem to
be the most accurate. As pointed out in the previous Section, in this
approach the Curie temperatures have been identified from the maxima of the
specific heat, according to eq.\,(13).

In Fig.\,3 the specific heat curves per lattice site for finite
clusters are illustrated vs. temperature for various concentrations $p$.
The temperatures corresponding to the maxima of these curves are denoted by
the circular markers in Fig.\,2. Although the location of the specific heat
peaks can be determined numerically from the curves, we found that the
application of the analytical formula (13) leads to more precise results.
It should be noted that the specific heat in Fig.\,3 behaves correctly from
the thermodynamic point of view, both in the low and high temperature
limits.

\section{Conclusions}

In the paper four approximate methods have been applied in order to study
the low dimensional PT Ising ferromagnet with selective dilution. Within
those methods the formulas for the Curie temperatures have been obtained.
The phase diagram has been calculated for various values of the
concentration parameter $p$ and the results of the different methods have
been compared.

The most accurate method is the one based on the exact numerical
calculations for finite clusters with the periodic boundary conditions. It
is worth noticing that within this method all thermodynamic properties can
be simultaneously calculated in the same computational time. However,
regarding the determination of the Curie temperature, this method is
applicable to the systems in which the maximum of the specific heat is
unambiguously related to the phase transition temperature. This is not
always the case; for instance, for the frustrated systems or even some
paramagnets where the so-called Schottky maximum is observed. Therefore, the analytical methods, giving better physical
insight and proper interpretation of the numerical results, are still
important in such studies.

Among analytical methods the PA approach can be especially recommended,
for it enables the self-consistent studies of all thermodynamic properties
based on the Gibbs potential. It also gives satisfactory accuracy when
compared with other approaches. As shown recently, this method can be
applied to the Heisenberg systems as well [24, 25]. It should also be noted
that for quantum systems the numerical diagonalization of the finite
cluster Hamiltonian is much less efficient than ECFC for the classical
Ising model.

For the model in question the antiferromagnetic interactions can also be considered. Then, for selective dilution $p > 0$ the frustrations will occur and the theoretical description becomes more complex. This problem should be a subject for separate paper.

\begin{acknowledgments}
The numerical calculations have been performed on the computer cluster
HUGO at the P.\,J.\,\v{S}af\'arik University in Ko\v{s}ice.

\end{acknowledgments}

\end{document}